\documentclass[3p]{elsarticle} 

\usepackage{lineno,hyperref}
\modulolinenumbers[1]

\journal{xxx}

\usepackage{amssymb}
\usepackage{setspace}
\usepackage{rotating}
\usepackage{multicol}
\usepackage{cancel}
\usepackage[usenames,dvipsnames]{color}
\usepackage{xcolor}
\usepackage{soul}
\soulregister\cite7
\soulregister\ref7
\soulregister\pageref7
\sethlcolor{limegreen}
\usepackage{amsmath}
\usepackage{lscape}
\usepackage{multicol}
\usepackage{multirow}
\usepackage{footnote}
\usepackage{url}
\usepackage{array,longtable}
\usepackage{colortbl}

\usepackage{rotating}
\usepackage{subcaption}
\usepackage[english]{babel}
\usepackage[labelsep=period,singlelinecheck=no]{caption}
\usepackage{enumerate}
\usepackage{epstopdf}
\usepackage{setspace}
\usepackage{dblfloatfix}
\usepackage{float}
\usepackage{flafter}

\usepackage{amsthm}
\usepackage{fancyhdr}
\usepackage{enumitem}
\usepackage{xfrac}
\usepackage{mdwlist} 
\usepackage{array}
\usepackage{tabularx}
\usepackage[mmddyyyy,12hr]{datetime}
\allowdisplaybreaks[1]

\usepackage{pifont}
\usepackage{wasysym}
\usepackage{url}

\usepackage{tikz}
\usetikzlibrary{calc}
\usepackage{ulem}
\tikzset{>=stealth}

\newcommand{\deriv}[2]{\genfrac{}{}{}{}{{\rm d}{#1}}{{\rm d}{#2}}}

\newcommand{\ds}{\displaystyle}
\renewcommand{\leq}{\leqslant}
\renewcommand{\geq}{\geqslant}
\newcommand\super[1]{$^{\text{#1}}$}

\newcommand*{\colorboxedAux}[3]{%
  \begingroup
    \colorlet{cb@saved}{.}%
    \color#1{#2}%
    \boxed{%
      \color{cb@saved}%
      #3%
    }%
  \endgroup
}
\newcommand*{\colorboxed}{}
\def\colorboxed#1#{%
  \colorboxedAux{#1}%
}

\newcommand{\sboxed}[2]{{\setlength\fboxsep{2pt}\colorboxed{#1}{#2}}}
\newcommand{\dboxed}[2]{{\setlength\fboxsep{1pt}\colorboxed{#1}{\colorboxed{#1}{#2}}}}

\definecolor{purp}{rgb}{0.5411765 0.1686275 0.8862745}
\definecolor{lightpurp}{rgb}{0.5 0.4 0.8}

\newcommand{\pfour}{P$_\text{4}$}
\newcommand{\etwo}{E$_\text{2}$}

\floatstyle{ruled}
\newfloat{algorithm}{tbhp}{lop}
\floatname{algorithm}{Algorithm}

\graphicspath{{figures/}}
\allowdisplaybreaks

\begin{document}

\begin{frontmatter}

\title{Reduced model for female endocrine dynamics: Validation and functional variations}

\author[ejg]{Erica J. Graham\corref{mycorref}}
\address[ejg]{Mathematics Department, Bryn Mawr College, Bryn Mawr, PA 19010, USA}
\cortext[mycorref]{Corresponding author. Address: Mathematics Department, 101 N. Merion Ave., Bryn Mawr, PA 19010, USA.}
\ead{ejgraham@brynmawr.edu}

\author[neaddress]{No\'emie Elhadad}
\address[neaddress]{Department of Biomedical Informatics, Columbia University, New York, NY 10032, USA}

\author[daaddress]{David Albers}
\address[daaddress]{Pediatrics Department, University of Colorado Denver--Anschutz Medical Campus, Aurora, CO 80045, USA}



%
\begin{abstract}
A normally functioning menstrual cycle requires significant crosstalk between hormones originating in ovarian and brain tissues. Reproductive hormone dysregulation may cause abnormal function and sometimes infertility. The inherent complexity in this endocrine system is a challenge to identifying mechanisms of cycle disruption, particularly given the large number of unknown parameters in existing mathematical models.  We develop a new endocrine model to limit model complexity and use simulated distributions of unknown parameters for model analysis. By employing a comprehensive model evaluation, we identify a collection of mechanisms that differentiate normal and abnormal phenotypes. We also discover an intermediate phenotype--displaying relatively normal hormone levels and cycle dynamics--that is grouped statistically with the irregular phenotype. Results provide insight into how clinical symptoms associated with ovulatory disruption may not be detected through hormone measurements alone.
\end{abstract}

\end{frontmatter}


\section{Introduction}

Female endocrine physiology is relatively poorly understood and is related to several high-impact diseases---including polycystic ovary syndrome (PCOS), endometriosis, and diabetes-related illnesses like non-alcoholic fatty liver disease---along with general reproductive health and mental health. Here we are primarily motivated by two of these: PCOS, a condition where menstrual periods are infrequent or prolonged and/or where excess male hormone (androgen) levels are present, and endometriosis, a condition where uterine-lining tissue grows throughout the body. Both conditions lead to severe complications and reduced quality of life. The systems physiologic understanding of such conditions are poorly understood, and few data and few models exist to explain their origin or to provide a basis for high-fidelity phenotypes. 

We want to forge the path for better understanding of female endocrine physiology anchored with mathematical physiology. The pathway for using quantitative descriptions of physiology to impact human health is non-trivial: (1) physiologic systems are complex; (2) mathematical models of physiology do not rely on first principles, but are rather idealizations of how we imagine the physiologic subsystems to function; (3) model verification is also complex, requiring data and a means of evaluating the model's ability to represent those data; (4) models and parameters must be personalized and synchronized with patient data; and, (5) translation of the model-based information is difficult and complicated. 

Within clinical practice, the most common usage of physiology is for reasoning by analogy or via qualitative relationships. For example, in an ICU we inject insulin to reduce dangerously high glucose levels, and this is done incrementally with a dose chosen to be small enough to avoid hypoglycemia according to protocols developed using qualitative knowledge of physiology and clinical trials. Then, when interventions are implemented such as glucocorticoid administration, clinicians use qualitative knowledge of physiology and pharamcokinetics to alter insulin doses using a trial-and-error,  incremental approach. This situation generalized across much of biomedicine, and we want to move beyond this reason-by-analogy approach. However, we face many roadblocks to move from qualitative use of physiology to quantitative use of mathematical physiology \cite{jamia_da}. In addition to the complexity of clinical data \cite{jamia_phys_ehr}, other roadblocks include a lack of data for validation, no direct translation between complex quantitative physiologic information and the clinical setting, the subsequent lack of clinical entry points, and the translation of inferred physiological information to understandable, actionable clinical knowledge. Nevertheless, the problems we choose to tackle, along with the basic  construction and usage of related models, can be directed by the real potential to impact our understanding of health, especially as far as informed decision-making and plausible data collection are concerned.

Here we focus on the subsystem of the female endocrine system that is thought to control the ovulatory cycle. In the qualitative, physiologically mechanistic, mathematical context, descriptions of ovulatory dysfunction are complicated. First, the physiology is complex: there are multiple mechanisms, both in the brain and in the ovaries, that can alter ovulatory function. Second, and more problematic, we are extremely data-limited: even qualitatively, there is no systematic, comprehensive way to classify dysfunction because dynamical time scales are long (weeks to years), intra- and inter-personal variability is high, and observable manifestation of dysfunction can have many sources whose delineation can be difficult to resolve with data. The invasiveness of collection procedures adds to the limited data at our disposal.

The data that can exist include primary reproductive hormone measurements, which are useful for delineating broadly defined clinical abnormalities and quantifying generalized ovulatory states. For example, two prototypical data sets reported in the literature include pituitary and ovarian hormones collected daily over the course of a typical cycle \cite{McLachlan1990,Welt1999}; we use the data in \cite{McLachlan1990} in this paper. The challenge is these data provide only a partial view to more subtle abnormalities. For example, PCOS can result in the complete absence of, or sporadic, ovulation. But, distinguishing between mechanisms governing these two observable clinical manifestations is difficult because clinically feasible diagnostic tools rely on measurements taken either at a single time point or over the course of a few hours \cite{YenJaffe}. We would require data spanning multiple months in order to build a comprehensive hormone profile with any hope of revealing important reproductive features, especially in the absence of clearly identifiable ovulatory states.  Still more confounding are conditions such as endometriosis, which lack a clear etiology, yet are linked to circulating hormone levels \cite{Lode2017}.

 In the present context, it is important to note that a high-fidelity, data-driven, robust and expansive definition of normal ovulatory function does not currently exist.  {This makes defining `normal' and `dysfunctional' a complex task, as dysfunction is usually defined as a deviation from normal.  Because of this, we will adopt a narrow definition of normal and consequently limit our ability to discover different-from-normal phenotypes.}  This limitation is due to the lack of data; with more data, the methodology here could provide more phenotypic fidelity. Ideally, we seek an alternative to patterns in hormone dynamics to distinguish between ovulatory phenotypes, with the hope that identifying underlying mechanisms of dysfunction lies in our ability to connect clinical symptoms with mechanisms that may not be apparent in hormone measurements alone. Given the problems identified above, modeling and analysis at this level of detail is not possible. 

In this paper we do not attempt to overcome all problems between the development of a model and the use of the model to help improve human health at once, but rather focus on two. \emph{First}, we begin with a model and reduce it, which has two consequences: (1) reduction of identifiable pathologies by decreasing the number of states and unknown parameters; and, (2) induction of a physiologic hypothesis about what variables are important for representing the female endocrine system related to PCOS and endometriosis. The result is a new endocrine model of ovulation. \emph{Second}, we evaluate the model's ability to represent data, delineate the time-dependent differences between normal and abnormal cycles \emph{given a cycle length}, and we examine emergent phenotypes \cite{high_fid_pheno} through analysis of the parameter space. Although model evaluation is limited by data availability, we construct the evaluation methodology such that, given more data, a more powerful, direct evaluation will be immediately possible.

\subsection{Paper road map} In Section \ref{sec:modeling}, we review the Graham-Selgrade model of ovulatory dynamics \cite{Graham2017}, which is the modeling starting point.  We then reduce this model by removing testosterone to create a new model of ovulatory dynamics; this is our first result. In Section \ref{sec:methods}, we introduce the computational, evaluation, and data-related machinery we use to validate the model and examine its ability to resolve and differentiate our narrowly defined phenotypes. In Section \ref{sec:results}, we follow with the computational validation of the new model as well as our investigation of the model parameters to study clinical phenotypes.

\section{Ovulation Model and its Reduction} \label{sec:modeling}
We develop a new endocrine model to describe essential processes in ovulation. This new model is a reduction of a model developed by Graham and Selgrade that uses ordinary differential equations to describe the ovulatory cycle under the influence of elevated androgens, namely testosterone \cite{Graham2017}.  Testosterone is a major element in the Graham-Selgrade model because androgen excess, or \textit{hyperandrogenism}, and insulin resistance are frequently associated with ovulatory dysfunction related to PCOS.  Although PCOS is an important disorder with many open questions regarding its etiology {\cite{Caldwell2017}}, we presently aim to examine generalized ovulatory dysfunction, which may or may not stem from previously defined clinical phenotypes, e.g., PCOS. Moreover, while we reduce the model by eliminating testosterone as a state variable, testosterone remains an implicit variable in the model; we verify testosterone-mediated dysfunction in the course of validating the reduced model. To achieve this goal we work to reduce the model to limit the size of parameter and state spaces, thereby reducing the complexity of the analysis and the data required to resolve phenotypes. We choose to begin with the Graham-Selgrade model because we deem it more amenable to reductions in the parameter space while retaining normal ovulatory dynamics,  in contrast to other related models that are based on delay differential equations \cite{Schlosser2000,Clark2003,Hendrix2014}.

\subsection{The Graham-Selgrade model} 
The Graham-Selgrade model \cite{Graham2017} is divided into three major subsystems: pituitary regulation, follicle dynamics, and ovarian steroidogenesis. Collectively, the model consists of 12 state variables, tracking serum concentrations of five important reproductive hormones, follicle stimulating hormone (FSH), luteinizing hormone (LH), estradiol (\etwo), progesterone (\pfour), and testosterone (T), along with precursors/intermediaries of LH, FSH, and T.  It also describes the dynamics of three follicular stages and of the follicle response to LH, termed \textit{LH sensitivity}. The final model contains 41\footnote{The model presented in \cite{Graham2017} contains a typographical error in one of the equations, which omits one parameter ($c_{\Phi,T}$) from the total parameter count cited.} unknown parameters which are estimated---{to a locally minimizing set}---by fitting the model to data from the literature \cite{McLachlan1990, Keefe2014}. The complete list of equations for the original Graham-Selgrade model may be found in  \ref{appendix:orig}.

\subsubsection{Compartmental model description}
The Graham-Selgrade model uses a compartmental framework to examine changes in ovulation due to increased androgens. The model follows the approaches of \cite{Schlosser2000,Clark2003,Hendrix2014} and comprises three major subsystems, which describe changes in the pituitary-ovarian axis with mechanisms of steroidogenesis. 
\begin{enumerate}[label=\Roman*.]
\item \textit{Pituitary regulation.} 
LH and FSH are the primary hormones produced by the pituitary gland. Synthesis and release of these hormones are regulated by ovarian steroid hormones, including  \etwo, \pfour, and  T. The equations governing changes in FSH and LH are split between releasable (denoted $FSH_\rho$ and $LH_\rho$) and serum (denoted $FSH$ and $LH$) pools of the hormones and incorporate stimulatory and inhibitory feedback by ovarian steroids. Using this compartmental approach, we can differentiate feedback processes governing pituitary hormone synthesis versus release.

Here we provide a generalized description of pituitary dynamics. Let $H(t)$ denote the serum concentration of a pituitary hormone (either FSH or LH) and $H_\rho(t)$ its releasable amount at time $t$. For $H=FSH, ~LH$, the differential equations governing releasable and serum quantities have the form
\begin{alignat}{4}
\deriv{H_\rho}{t} &= k_\text{synthesis}(\cdot) - k_\text{release}(E_2,P_4) H_\rho, \label{eq:hrho}\\
\deriv{H}{t} &= k_\text{release}(E_2,P_4) H_\rho/V - \delta_H H \label{eq:hgen}.
\end{alignat}
Each $k(\cdot)$ term denotes a function of state variables and describes the change in hormone levels due to the process indicated. Synthesis of FSH and LH is determined by different processes---with precise arguments to $k_\text{synthesis}$ omitted to reflect this---whereas their release is mediated solely by \etwo\ and \pfour. Release into the serum is scaled by the blood volume, $V$, and clearance of the hormones is assumed to be a first-order process, with rate constant $\delta_H$.  Regardless of the highly nonlinear form of ovarian feedback, the subsystem remains linear in $H_\rho$ and $H$. Collectively, the pituitary subsystem comprises four differential equations, with Equations \eqref{eq:hrho} and \eqref{eq:hgen} defined explicitly for both FSH and LH.  

\item \textit{Follicle dynamics.} 
Follicle growth, maturation, and differentiation are assumed to occur in a series of three sequential stages: (1) \textit{follicular}, (2) \textit{ovulatory}, and (3) \textit{luteal}. We denote these using variables $\Phi(t),~\Omega(t)$, and $\Lambda(t)$, respectively.  The  follicular phase is characterized by recruitment and growth of stimulated follicles.  The ovulatory phase is characterized by ovum release from a designated follicle in response to a mid-cycle surge in LH. Finally, the luteal phase is characterized by the formation and, in the absence of fertilization, regression of the corpus luteum. The three follicular stages are modeled as follows:
\begin{alignat}{4}
\deriv{\Phi}{t} =&\,  k_\text{recruitment}(T) + k_\text{growth}(FSH,T) \Phi - k_\text{ovulation}(FSH,LH) \Phi,\\
\deriv{\Omega}{t} =&\, 
		k_\text{ovulation}(FSH,LH) \Phi -  
		k_\text{luteal}(S) \Omega, \\
\deriv{\Lambda}{t} =&\, k_\text{luteal}(S) \Omega - k_\text{regression}(S)\Lambda. 
\end{alignat}
Transitions to subsequent stages are unidirectional and depend on pituitary hormone levels. The model also incorporates a role for T in follicle recruitment and growth. Graham and Selgrade further define a new LH support variable, $S(t)$, to model the tonic LH-dependence of growth and premature regression of the corpus luteum. Specifically, $S$  decays exponentially (with rate $\delta_S$) to 0 in the absence of LH and approaches a maximal level of 1 for sufficiently large LH: 
\begin{equation}
\deriv{S}{t} =  k_\text{activation}(LH)(1-S) - \delta_SS.
\end{equation}

\item \textit{Ovarian steroidogenesis.} Throughout the ovulatory cycle, follicles may produce \etwo, \pfour, and T. Intracellular steroid production is primarily FSH- and LH-dependent during a typical cycle and is subject to functional maturation of individual follicles. This subsystem exploits the \textit{two-cell two-gonadotropin} theory of ovarian steroid production, which describes the differential functionality of theca cells and granulosa cells within ovarian follicles \cite{YenJaffe}.   The Graham-Selgrade model also introduces a semi-mechanistic description of testosterone production for examining a role for insulin in promoting hyperandrogenism. For $T_\gamma(t)$ denoting the `intermediate' concentration of T destined to be converted into \etwo, we write 
\begin{alignat}{4}
\deriv{T_\gamma}{t} =&\, k_\text{entry}(LH,\alpha) - k_\text{aromatization}(FSH)T_\gamma. \label{eq:Tgorig}
\end{alignat}
In a growing follicle, theca cells compose the outermost layers of cells surrounding the ovum and granulosa cells the innermost layers. Importantly, theca cells possess androgen (i.e. T) production machinery and are stimulated by LH alone, whereas only neighboring granulosa cells can convert these androgens into estrogens, in an FSH-dependent process called \textit{aromatization}. Therefore, we consider $T_\gamma$ to reflect the average concentration of T that enters granulosa cells from  theca cells. 

Finally, we model the major ovarian outputs of the model: serum concentrations of \etwo, T, and \pfour:
\begin{alignat}{4}
\deriv{E_2}{t} =&\,  k_\text{basal,E} - \delta_E E_2 + k_\text{aromatization}(FSH)T_\gamma \cdot  f_E(\Phi,\Omega,\Lambda), \label{eq:E2orig}\\
\deriv{T}{t}~ =&\,  k_\text{basal,T} - \delta_T T + \left[k_\text{\parbox{1.2cm}{ovarian\\[-6pt]production}}(LH,\alpha) + k_\text{\parbox{1.3cm}{peripheral\\[-6pt]production}}(LH,\alpha)\right]\cdot f_T(\Phi,\Omega,\Lambda), \label{eq:Torig}\\
\deriv{P_4}{t} =&\,  k_\text{basal,P} - \delta_P P_4  + k_\text{secretion} (LH) \cdot  f_P(\Phi,\Omega,\Lambda). \label{eq:P4orig}
\end{alignat}
The first two terms in Equations \eqref{eq:E2orig}--\eqref{eq:P4orig} represent  basal secretion by the adrenal gland and first-order clearance of individual steroids,  defined by rate constants $k_\text{basal,I}$ and  $\delta_I$, respectively, where $I = E,T,P$. The last term in each equation defines secretion of steroid hormones into the circulation, which is assumed to occur immediately upon production. The average production rate per follicle is multiplied by a function $f_I(\Phi,\Omega,\Lambda), ~I = E,T,P$, that describes the relative contribution of each follicular stage to the production of a given steroid.

Importantly, steroidogenesis is altered through feedback from FSH and LH, according to the two cell-two gonadotropin theory. Whereas LH is required almost exclusively for T (theca only) and \pfour\ (theca and granulosa) production, FSH is entirely responsible for \etwo\ (granulosa only). Because \pfour\ is an androgen precursor in the theca, it is assumed that circulating \pfour\ is produced primarily by granulosa cells for modeling purposes. To address insulin's influence in ovulatory dysfunction, the Graham-Selgrade model contains a detailed formulation of T production, wherein ovarian and peripheral conversion of T from its precursors are treated as two distinct processes. In Equations \eqref{eq:Tgorig} and  \eqref{eq:Torig}, the parameter $\alpha$ represents the relative degree to which insulin may increase T production.   
\end{enumerate}

\subsection{A new model of endocrine regulation:  Graham-Selgrade model reduction} \label{sec:reduction}

A natural course of action in determining salient model behavior is global sensitivity analysis (GSA) of parameters. This allows us to determine the relative sensitivity of model output to changes in the parameters. There are multiple challenges associated with the Graham-Selgrade model that make GSA a suboptimal next step in model analysis. First, the model contains considerably more parameters than the data available for estimation. Second,  coupling between state variables is highly nonlinear. Third, stable limit cycles are not guaranteed for all parameter combinations. Collectively, standard GSA approaches provide limited insight. In particular, a PRCC-based approached would be inappropriate, as simulations do not yield monotonic hormone responses that can be interpreted in any meaningful way (preliminary work, not shown). Alternatives such as the extended Fourier amplitude sensitivity test (eFAST) may also prove more useful, as discussed in \cite{Marino2008}. However, selection of appropriate model output remains a challenge.  In light of these observations, we hypothesize that structural reduction of the model may provide greater insight to relevant and essential processes governing the typical ovulatory cycle.  We may also use the resulting framework to examine more general questions of ovulatory function separate from the pathologies associated with PCOS. Here we  introduce the major modifications to effectively reduce the number of unknown model parameters.

\subsubsection{Removal of testosterone}
Complete data sets that track the pituitary hormones, LH and FSH, as well as ovarian steroids \etwo\ and \pfour\ during the course of an entire cycle are uncommon but not completely absent. What is missing is a complete hormone profile that also includes androgen levels through the course of a normal ovulatory cycle. Based on this information, we eliminate T (and hence insulin) entirely from the Graham-Selgrade model. With this adjustment, we remove Equation \eqref{eq:Torig} entirely and set $\alpha=0$ and $T(t) = T_0$ for all $t$.  We then adjust the remaining differential equations as needed to eliminate  $T$ (and $T_\gamma$) coupling. 

\paragraph{Effect on steroidogenesis}  We assume that intermediate T transferred from theca to granulosa cells is immediately converted into \etwo. Then using a similar reduction approach in \cite{Graham2017} applied to a more mechanistic steroidogenesis model, we set $\dot{T_\gamma} =0$ and solve for $T_\gamma$. Substituting the resulting expression  and the original parameters into Equation \eqref{eq:E2orig} gives
$
\dot E_2 =  e_0 - \delta_E E_2 + t_{g1}F_1(LH) \cdot (\Phi + \eta \Lambda S),
$
where $F_1(LH) = {LH^2}/[{\kappa_1 LH^2 + \kappa_2 LH + \kappa_3}]$. We further observe that for sufficiently large $LH$, $\kappa_3 \ll \kappa_1 LH^2 + \kappa_2 LH$, so we redefine $F_1 (LH) = LH/(LH+\kappa_2)$ and rescale parameters accordingly (see Equation \ref{eq:e2}). 

Notably, the reduced version of $\dot E_2$ removes the FSH-dependence on \etwo\ production from the original framework. For a normally ovulating cycle, we may then consider FSH to be permissive for \etwo\ synthesis within granulosa cells. Further, the time scale on which FSH alters follicular expression of aromatase---via the steroidogenic acute regulatory protein---is significantly shorter ($\sim$ minutes) than that of a typical cycle  ($\sim$ days) {\cite{Miller2011}}. As such, we assume any tonic level of FSH allows for proper steroid synthesis, and the degree to which this occurs depends solely on the functional maturation of follicles in the reduced model. Therefore, we relegate FSH-dependent \etwo\ production to a nonessential process and focus instead on other sources of pathological behavior within the cycle.


\paragraph{Effect on pituitary regulation} In the original model, the influence of T is restricted to synthesis of releasable LH: T is assumed to increase basal LH production and to prevent \pfour-mediated inhibition of all LH production. To eliminate T from the necessary terms, we redefine the affected parameters given our assumption of relatively constant T levels. In particular, we set $v_{0L}^\text{new} = v_{0L}^\text{old} T_0/(K_{L,T}+T_0) $ and $K_{iL,P}^\text{new} = K_{iL,P}^\text{old}(1+c_{L,T} T_0)$ (see Equation \ref{eq:rlh}). 

\paragraph{Effect on follicle dynamics}  In the original model, T serves two functions: (1) it influences the rate at which very immature follicles enter follicular ($\Phi$) stage to begin gonadotropin-dependent growth and differentiation, and (2) it increases follicle sensitivity to FSH signaling. To eliminate T from this subsystem, we note that the originally estimated basal rate of T-mediated  follicle recruitment is $f_0\sim \mathcal O(10^{-3})$. It is therefore unlikely that this process contributes substantially to the function of a normal cycle, and so we  simply set $f_0=0$. Following our approach  in the pituitary subsystem, we redefine the FSH sensitivity parameter to be $h_1^\text{new} = h_1^\text{old}/(1+c_{\Phi,T})$, provided that follicular FSH sensitivity remains constant throughout the menstrual cycle (Equation \ref{eq:phi}).

\subsubsection{Simplifying assumptions} \label{sec:simplify}
In addition to eliminating T as a state variable, we make two simplifying assumptions to further reduce the number of unknown parameters. These parameters are chosen in consideration of the important biological components that must be maintained in order to consider the resulting model an accurate and useful representation of the menstrual cycle under physiological conditions. Rather than focusing on a topologically equivalent system, we focus on preserving plausible biological mechanisms.

\paragraph{FSH-dependent LH sensitivity} An essential event in ovulation is the upregulation of LH receptors in the late follicular stage  (stage $\Phi$). This is an FSH-dependent process occurring within sufficiently mature follicles. The Graham-Selgrade model assumes FSH increases follicle sensitivity to LH during stage $\Phi$. In the model reduction, we assume the maximal sensitivity parameter, $h_2$ remains constant. This is a reasonable simplification, as $h_2 \sim \mathcal O({10^{3}})$ and $c_{\Phi,F}\cdot {\ds \max_t}\{FSH(t)\} \sim \mathcal O(1)$ in the original model.

\paragraph{LH-dependent \pfour\ production} \pfour\ conversion within theca and granulosa cells requires enzymes that are regulated by LH. The estimate for the half-maximal LH concentration, $h_p$, that stimulates \pfour\ production in the original model is lower than the simulated LH concentration during the luteal phase (where \pfour\ attains peak concentration). Therefore, we assume that the steroid production per follicle is constant at the maximal rate $p$ in LH.

\subsection{Reduced mathematical model equations} \label{sec:reducedEq}
The new model is given by Equations \ref{eq:rfsh}--\ref{eq:p4} and contains 10 differential equations.  With 27 unknown parameters, we have reduced the parameter space by more than a third. Terms in the model that have been altered due to removal of testosterone are boxed with a single line. Those that result from additional simplifying assumptions as described in Section \ref{sec:simplify} are boxed with a double line. For comparison, the original model equations are listed in \ref{appendix:orig}.
\xdef\xlab{1.1in}
\begin{flalign}
\parbox{\xlab}{\small Releasable FSH:}&&&
	\deriv{FSH_{\rho}}{t}\hspace{-6em} &=\,& \frac{v_F}{1+c_{F,I}\frac{S\Lambda}{K_{iF,I} + S\Lambda}} - k_{F} \frac{1+c_{F,P}P_4}{1+c_{F,E}E_2^2}FSH_{\rho} \label{eq:rfsh} \\
\parbox{\xlab}{\small Serum FSH:}&&&
	\deriv{FSH}{t} \hspace{-6em} &=\,& \frac{1}{V} \cdot k_{F} \frac{1+c_{F,P}P_4}{1+c_{F,E}E_2^2}FSH_{\rho} - \delta_{F} FSH \label{eq:fsh}\\
\parbox{\xlab}{\small Releasable LH:}&&&
	\deriv{LH_{\rho}}{t} &=\,& \left[\sboxed{black}{v_{0L}}
	 +  \frac{v_{1L}E_2^n}{K_{mL}^n + E_2^n}\right] \cdot \label{eq:rlh} 
	 \frac{1}{1+\sboxed{black}{P_4/K_{iL,P}}} - k_{L} \frac{1+c_{L,P} P_4}{1+c_{L,E}E_2}LH_{\rho}   \\
\parbox{\xlab}{\small Serum LH:}&&&
	\deriv{LH}{t} &=\,& \frac{1}{V} \cdot k_{L} \frac{1+c_{L,P} P_4}{1+c_{L,E}E_2}LH_{\rho} - \delta_{L} LH \label{eq:lh}\\
\parbox{\xlab}{\small Follicular phase:}&&&
\deriv{\Phi}{t} &=\,&  \left(\frac{f_{1} FSH^2}{\sboxed{black}{{h_1^2}} +FSH^2} 
	 -	\frac{f_2LH^2}{\dboxed{black}{h_2^2} + LH^2} \right) \cdot \Phi \label{eq:phi} \\
\parbox{\xlab}{\small Ovulatory phase:}&&&	 
\deriv{\Omega}{t} &=\,& 
		\frac{f_2LH^2}{\dboxed{black}{h_2^2} + LH^2} \cdot \Phi - 
		w   S   \Omega \label{eq:omega} \\
\parbox{\xlab}{\small Luteal phase:}&&&
\deriv{\Lambda}{t} &=\,& w S  \Omega - l  (1-{S})\Lambda  
  \label{eq:lambda} \\
\parbox{\xlab}{\small LH support:}&&&
\deriv{S}{t} &=\,& \hat s \frac{LH^4}{LH^4+h_s^4}(1-S) - \delta_S S \label{eq:s}\\
\parbox{\xlab}{\small Serum \etwo:}&&&
\deriv{E_2}{t} &=\,& e_0 - \delta_E E_2 + \sboxed{black}{t_{g 1} \frac{LH}{LH+\kappa_2}}   \cdot 
		{(\Phi + \eta \Lambda  S)} \label{eq:e2}\hspace{3.2in}\\
\parbox{\xlab}{\small Serum \pfour:}&&&
\deriv{P_4}{t} &=\,&  - \delta_P P_4  + \dboxed{black}{p} { \Lambda  S  } \label{eq:p4}\hfill
\end{flalign}

\section{Computational Methods and Model Evaluation}\label{sec:methods}

\subsection{Terminology: physiological vs. mathematical cycles}
To discuss model evaluation and results, we explicitly distinguish between physiological and mathematical notions of a `cycle'. When discussing properties of mathematical ovulation, we explicitly refer to the \emph{inter-ovulatory interval} (IOI), which denotes the length of time between consecutive simulated LH surges. Physiologically, the IOI is equivalent to the time between two \textit{ovulatory cycles}; however, multiple IOIs may be required before the solution completes a single {mathematical (limit) cycle}. For clarity and consistency, we restrict our generalized use of `cycle' to refer to physiological ovulation and IOI to the calculated times between these cycles.

\subsection{Data}
We use two data sets, one synthetic and one real.  The first data set, the synthetic data set, is generated using the original model \cite{Graham2017}. This data set is used to show that the reduced model captures most of the dynamics of the original model. {Effectively, we also show that explicit inclusion of testosterone is not needed to capture important features of the physiological--but not necessarily pathological--hormone dynamics.} The second data set is hormone data available in \cite{McLachlan1990}. These data contain \emph{average} daily measurements for 33 normally cycling women during the course of one complete ovulatory cycle for FSH, LH, \etwo, and \pfour. This second data set is used to demonstrate the both ability of the model to estimate data well, and how to use the model to better understand physiology, given data. 

\subsubsection{Limitations and complexities of data and analytical challenges}

The available data have three primary limitations that influence our work. First, recall that normal is generally poorly defined, where `normal' means no known pathophysiologic cycle features. Second, it is known that there is substantial variation in IOIs even for an individual.  For example, it is not uncommon for the same person to have IOIs that vary from 20 to 40 days; these data obscure such intraindividual variability by taking an average.  And third, because the data are an average, they induce three potential issues whose presence we may not be able to detect: (i) an average can fail to represent anyone if the mean is not representative of the population, (ii) an average smooths variability observed personal \emph{daily} variability, variability that can be substantial, is not present in data and will not be explicitly estimated by the models, and (iii) variability of cycle length and dynamics coupled to cycle length for both ill-defined normal cycle length and abnormal cycles is entirely missing.

These data limitations impact the analysis in a fundamental way, most notably, the generalizability of our results.  We develop a phenotypic analysis subject to a standard IOI of 31 days and relative to average hormone dynamics.  It is surely possible that the average  data do not represent individuals well. Moreover, it is surely possible that hormone dynamics of different IOIs will be different.  Finally, within the model, there are two ways of generating variable IOIs: (1) one can change parameters that alter a constant IOI, or (2) one may define parameter regimes that allow for variable IOIs between consecutive ovulatory events.  Given that our data set is limited to a month and is an average, we cannot investigate the distinction between these two model-based parameter differences.  In short, the data set limits the generalizability of data-based model validation, but our demonstration for how this pathway will work in the future {remains} relevant. These limitations demonstrate the urgent need for the collection of a more realistic and expansive hormone data.

\subsection{Model comparison: the reduced model vs. the original model}

Equipped with computational machinery for estimating parameters given data, we want to demonstrate the capability of identifying and defining phenotypes that emerge from the model. As previously mentioned, we cannot move beyond what can be tested with data we have, but for our purposes this will not limit us.  We  focus on two model-defined phenotypes, physiologic and pathophysiologic, and codify these as {regular} or {irregular} cycle behavior, respectively. We assign these to cases of model-generated data by defining a set of attributes that may distinguish between physiological and pathological ovulatory function. Given the complex cross-talk in the reproductive endocrine system, \emph{analysis of hormone concentrations alone} likely provides insufficient insight into the subtleties of ovulatory dysfunction. To overcome this challenge, we proceed by identifying a collection of parameters giving rise to predetermined {regular} or {irregular} cycle behavior. To accomplish this, we implement an algorithm that allows us to carry out a comprehensive evaluation of the reduced model.

\subsubsection{An algorithm for comprehensive model evaluation}
We introduce an algorithm to compare the new endocrine model to the original Graham-Selgrade framework. In this section, we provide an overview (see Algorithm \ref{algorithm}) and a detailed description of our implementation of the five-step algorithm.
\begin{algorithm}[tb]
{
\begin{enumerate}[leftmargin=0.6in,label=\textbf{Step \arabic*}.,ref=\theenumi] \raggedright
    \item {Generate synthetic data set using Equations \eqref{eq:rfshOG}--\eqref{eq:p4OG}.} \label{synthetic}
 
    \item {Optimize reduced model parameters using weighted least squares and synthetic data.} \label{redfit}
    
    \item Run $N$ Monte Carlo simulations, initialized with perturbed best-fit parameters from Step \ref{redfit} and refit to clinical data. 
    \label{montecarlo}
    
    \item Compute numerical solutions for each parameter profile generated in Step \ref{montecarlo}, and store resulting hormone data over multiple cycles. \label{MCprofiles}
    
    \item Use results of Steps  \ref{montecarlo} and \ref{MCprofiles} to define salient phenotypes and distributions for each of the 27 reduced model parameters. \label{pardist}
    
\end{enumerate}
}
\caption{Comprehensive model evaluation.}
\label{algorithm}
\end{algorithm}

\begin{description}
\item[Step \ref{synthetic}: Synthetic data.] We generate the synthetic data by numerically solving the system \eqref{eq:rfshOG}--\eqref{eq:p4OG}, using the parameters in \cite{Graham2017}, for a sufficiently long time to approach a stable limit cycle for normal ovulation. We then align the trajectories so that the LH surge occurs at the end of day 15 of the first cycle. Finally, we extract daily data between days 0 and 30 and then, to avoid propagated numerical inaccuracies, repeat the cycle twice more for each variable. We also expand the set of data to include $\Phi$, $\Omega$, and $\Lambda$, under the assumption that follicular dynamics should follow a similar pattern to the original model. With the exclusion of T from the model, we have a total of 7 state variables (including FSH, LH, \etwo, and \pfour) with $n=93$ data points each. 

\item[Step \ref{redfit}: Optimization.] 

To determine how well the new model compares to the original model, we estimate the 27 parameters of the reduced model by fitting output to the synthetic data.  We capture essential cycle behavior with the optimized parameters using a weighted least squares approach. For a given variable $X_i(t)$, where $i \in \{ FSH, LH, E_2,P_4,\Phi, \Omega,\Lambda\}$, we first assign default weights $w_i = 1/{\rm Var}(X_i(t))$ to each data point at $t_j=0,1,\ldots,92$. Because we cannot guarantee the expected behavior of follicular dynamics, we do not incorporate additional time-dependent weights for $i=\Phi,\Omega,\Lambda$. However, for the hormones we increase weights by variable factors at important peaks, troughs, and plateaus within the data. These weights are adjusted to acquire the best qualitative fit to the data, with the understanding that local minimization of the cost function may be sensitive to variation in weights and may not produce a globally optimal solution.

Let  $\mathbf{y}_i$ represent the vector of measurements corresponding to reduced model output variable $\mathbf{x}_i(\boldsymbol\varphi)$, defined by parameters $\boldsymbol\varphi$.   We define the optimization problem that minimizes the sum of the squared error as
\begin{equation}
\min_{\boldsymbol \varphi} \frac{1}{|V| \cdot n}\sum_i w_i  ||\mathbf y_i - \mathbf x_i(\boldsymbol\varphi)||^2,  
\label{eq:cost}     
\end{equation}
where $V = \{FSH, LH, E_2,P_4,\Phi, \Omega,\Lambda\}$ and denote the optimal parameter vector satisfying Equation \eqref{eq:cost} by $\boldsymbol\varphi^*$. We use \textsc{Matlab}'s \texttt{fminsearch}, which implements the Nelder-Mead simplex method, to determine the optimal $\boldsymbol\varphi$. In most cases, initial parameter guesses are taken from the original model. In others, they are derived from the adjustments made in the reduction process, as described in Section \ref{sec:reduction}. The best-fit parameters are listed in Table \ref{tab:parREDUCED} in the Appendix.

\item[Step \ref{montecarlo}: Monte Carlo simulations.] We determine the distributions of the 27 model parameters using Monte Carlo simulations to generate a collection of best-fit parameters using various initial guesses in the estimation scheme described in Step \ref{redfit} and by comparing the output to data. We first assume that the  values in   $\boldsymbol \varphi ^*$  represent mean quantities and that initial guesses, $\boldsymbol \varphi^{(0)}$, are uniformly distributed within $\pm10\% $ of the mean. That is,  $\varphi^{(0)}_k \sim  U(0.9\varphi^*_k,1.1\varphi^*_k)$ for $k=1,2,\ldots,27$.   To ensure a representative sampling of $N$ parameter combinations from each individual subinterval of length $1/N$ ranging from $0.9\varphi^*_k$ to $1.1\varphi^*_k$, we use  Latin hypercube sampling (LHS) and randomly generate initial parameter guesses for the Monte Carlo simulations (see \cite{Blower1994} for further discussion on LHS). For each initial parameterization we minimize a cost function similar to Equation \eqref{eq:cost}, this time with measurements $\mathbf y_i,~i \in \{ FSH, LH, E_2,P_4\}$  taken from clinical data available in  \cite{McLachlan1990}. 

\item[Step \ref{MCprofiles}: Range of simulated model output.] We numerically solve the reduced model over 186 days using the estimated parameters and generate an ensemble of these solutions via Monte Carlo sampling. We align each LH surge (assuming one exists) to day 15 and determine the length of each IOI. The \textit{LH surge} is defined to be a peak LH concentration that is followed by an apparent luteal phase; any other local maxima in LH failing to meet this criterion are ignored.  Because we have restricted our sampling scheme in Step \ref{montecarlo}, we guarantee that the model does not approach a stable equilibrium. Although this limitation does not capture complete ovarian failure (i.e., the absence of a cycle at all), it does allow for reasonable comparisons in the presence of oscillatory dynamics.  

\item[Step \ref{pardist}: Phenotypes and parameter distributions.] ~
\begin{enumerate}[label=(\alph*)]
    \item \textit{\textbf{Defining phenotypes.}} We implement a two-step process for determining distinct phenotypes using model output. First, we use the values of extreme IOIs to ensure that the presence of abnormally long or short IOIs at any time is considered pathological. In particular, we assign a \textit{regular} phenotype to simulations resulting in both minimal and maximal IOIs between 25 and 35 days, which is the textbook standard range for normal ovulatory cycles \cite{YenJaffe}. We assign an \textit{irregular} phenotype to simulations failing to satisfy this criterion. Second, we compute the mean squared error (MSE) between the data (LH, FSH, \etwo, and \pfour) and each simulation. Then we use the minimal MSE attained by an irregular phenotype to define a threshold for secondary regular phenotypes: \textit{regular$^+$} refers to regular phenotypes with MSE \uline{strictly less than} the computed threshold, and \textit{regular$^-$} to regular phenotypes with MSE \uline{at or above} the computed threshold. \label{alg:pheno}
     
   \item  \textit{\textbf{Computing parameter distributions.}} We construct empirical parameter distributions based on the optimized parameter sets obtained from the Monte Carlo simulations.  First, we apply the primary (regular vs. irregular) phenotype classification criteria to the $N$ Monte Carlo samples. Then we normalize the population sizes of individual phenotypes by subsampling the associated parameter distributions at their respective frequencies to an arbitrarily chosen size of $2000$.  \label{alg:pardist}
 \end{enumerate}

\end{description}

\subsubsection{Statistical Methods}
The addition of a phenotype classification generates several interesting questions that may be explored with the use of statistical and probabilistic tools.
\paragraph{Two-sample Kolmogorov-Smirnov (KS) test} In the present work, we seek phenotypic differences determined by model parameters. The KS test is used to determine whether two samples are drawn from the same distribution \cite{Rohatgi2000, Mora2015}. The test uses the Kolmogorov-Smirnov statistic, which is defined as the $L_\infty$ norm of the distance between two cumulative probability distribution functions. For each parameter, we then apply  \texttt{ks.test}, the R implementation of the two-sampled KS test, to analyze the phenotype-specific empirical distributions generated from our simulations.

\paragraph{t-Distributed stochastic neighbor embedding (t-SNE)} Beyond the structure manually imposed on the Monte Carlo dataset, we are interested in determining whether distinct phenotypes can be identified in another way. Patterns in the generated data may depend on any of 93 data points for each of four hormones, or any of the 27 parameter estimates. Without a comprehensive understanding of the interplay between each of these elements, we seek a methodology that will answer the binary question of whether there are inherent differences (seen or unseen) between regular and irregular phenotypes. t-SNE  is a machine learning tool for reduction of high-dimensional data to lower dimensions \cite{vanderMaaten2008}. We use the \textit{Rtnse} package in R to apply the t-SNE and determine whether phenotypes can be clearly clustered by a profile of select model parameters.

\section{Computational Results}\label{sec:results}

\subsection{Comprehensive reduced model evaluation}
\begin{figure*}[tb]
\centering
\includegraphics[width=.975\textwidth]{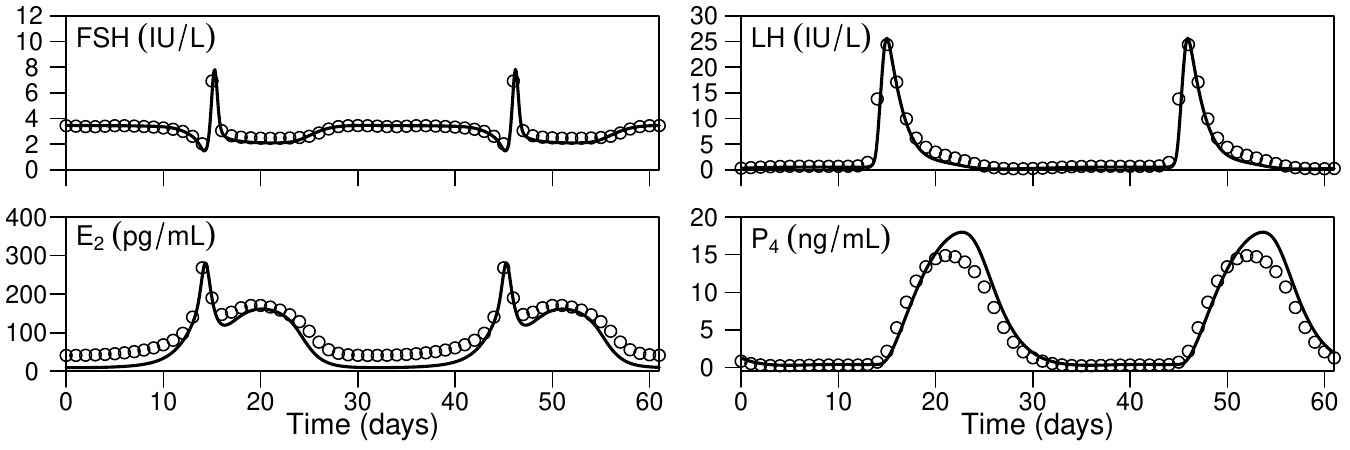}
\caption{Fit of reduced model to Graham and Selgrade model \cite{Graham2017} over 61 days. FSH and LH are displayed in standard international units according to the 2\super{nd} international reference preparation, where 1 IU FSH = 45 $\mu$g and 1 IU LH = 15 $\mu$g \cite{Labhart2012}. Conversion factors are based on the NIH preparation used in \cite{McLachlan1990}.}
\label{fig:comparison}
\end{figure*}
\subsubsection{Generalized model behavior}
Following Steps 1 and 2 of Algorithm \ref{algorithm}, we simulate the reduced model and compare results to the original model. We then use the parameterized model to simulate testosterone-mediated dysfunction, as further verification that the reduction is a plausible replacement of the original system.
\paragraph{Qualitative features} In Figure \ref{fig:comparison} we numerically solve the reduced model using the best-fit parameters and compare the result to output from the original Graham-Selgrade model. The qualitative dynamics are well captured, with the primary quantitative discrepancy related to $P_4$. This arises due to an overshoot of the data in the luteal stage $\Lambda$  during the mid-luteal stage (roughly 3--7 days after the simulated preovulatory LH surge, not shown). Since \pfour\ levels are known to peak clinically around this time, we consider this behavior to be within a physiologically relevant and normal range for the hormone. Further, because we assume that the ovarian stages are crude approximations to actual follicular dynamics, there may be substantial variability in the trajectories that may nevertheless yield normal ovulatory function, as illustrated in \cite{Keefe2014}.

\paragraph{Verification of testosterone-mediated dysfunction} A fundamental change in the reduced framework is the omission of testosterone, T. Although absent from the model, we may still examine how T might influence pathological ovulation. This approach also serves as proof of concept when using the reduced model in lieu of the original one. 

To re-incorporate T into the present framework, we modify relevant parameters. Following \cite{Graham2017}, we let $\alpha$ denote the degree of insulin influence, where $\alpha=0$ reflects a normal state with basal insulin (and hence T) levels. Assuming testosterone remains constant over time, we define its concentration using a linear function in $\alpha$, denoted $T_\alpha$:
\begin{equation}
T_\alpha = T_0 \cdot [ 1+ (\delta_T-1)\cdot (1+\alpha)]/\delta_T,
\label{eq:Talpha}
\end{equation}
where $T_0$ is the initial T concentration in the absence of hyperinsulinemia and $\delta_T$ is the first-order clearance rate of T from the blood, as defined originally. The parameters to be altered by T in the reduced model are $v_{0L}$, $K_{L,P}$, and $h_1$. We only consider the case of normal luteinization (see \cite{Graham2017} for details) because we have omitted FSH-dependent upregulation of follicle LH receptors, which would impact parameter $h_2$. To incorporate the necessary modifications to the current model, we redefine the parameters $v_{0L} \to v_{0L} \xi _1$, $K_{L,P} \to K_{L,P} \xi_2$, and $h_1 \to h_1\xi_3$ for $\alpha>0$, where
\begin{subequations}
\begin{align}
\xi_1 &= \frac{(\beta_1 + T_0) \cdot T_\alpha }{(\beta_1+T_\alpha)\cdot T_0},\\
\xi_2 &= \frac{1+\beta_2 T_\alpha}{1+\beta_2 T_0}, \text{ and}\\
\xi_3 &= \frac{1+\beta_3}{1+\beta_3 T_\alpha/T_0}.
\end{align}\label{eq:xi}
\end{subequations}%
\begin{figure*}[btp]
\begin{center}
\begin{minipage}[t]{.45\textwidth}\centering
\includegraphics[height=2in,trim=.8cm 0in 0in 0in,clip=true]{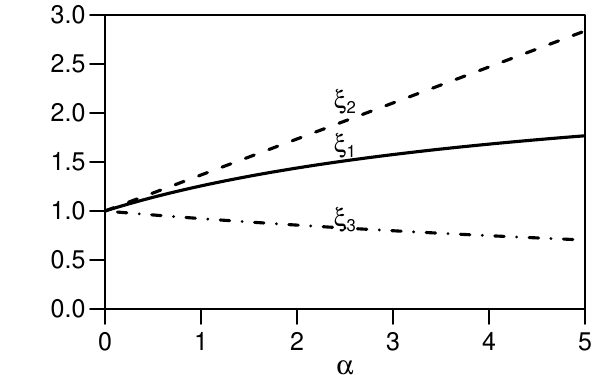}
\caption{Dimensionless functional forms used to incorporate T into reduced model, as in Equations \eqref{eq:Talpha} and \eqref{eq:xi}. Each $\xi_i$ contributes a T-dependent change (percent increase or decrease) in relevant parameters from the original model \cite{Graham2017}. $\xi_1$ increases LH synthesis parameter $v_{0L}$, $\xi_2$ increases \pfour-mediated LH inhibition parameter $K_{iL,P}$, and  $\xi_3$ decreases FSH sensitivity parameter $h_1$. $\alpha$: degree of insulin influence.}
\label{fig:xi}
\end{minipage} \hfill \begin{minipage}[t]{.5\textwidth} \centering
\includegraphics[height=2in]{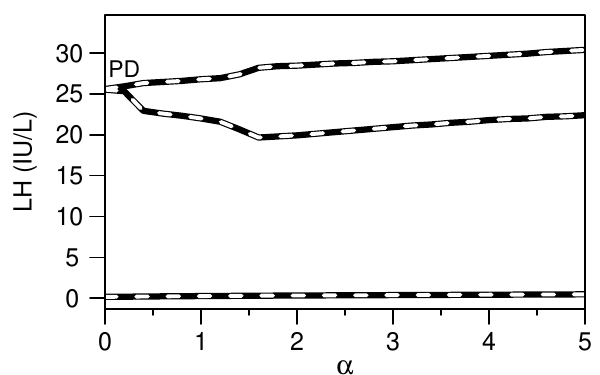}
\caption{Simulated bifurcation diagram depicting adjusted role for T and insulin influence ($\alpha$). Maximal and minimal LH concentrations are shown for various values of $\alpha \geq 0$. For $\alpha < 0.2$, LH oscillates between two values, suggesting a stable limit cycle. LH peaks alternate between consecutive IOIs for $\alpha \geq 0.2$, suggesting a period-doubling bifurcation (PD) with stable oscillations. 
}
\label{fig:tstlh}
\end{minipage}
\end{center}
\end{figure*}
The $\xi_i$ in Equations \eqref{eq:xi} determine the scaling of the model parameters as insulin influence increases and are plotted in Figure \ref{fig:xi}. The constants $\beta_i$ are defined according to the original model, with the caveat that bifurcation values of $\alpha$ may be shifted based on the values of these parameters. The derivation of the $\xi_i$ are given in \ref{appendix:tst}.

In Figure \ref{fig:tstlh}, we plot the long-term local maximum and minimum values corresponding to the LH surge for $0\leq\alpha \leq 5$. A stable limit cycle is roughly evident for $\alpha < 0.2$, with an  apparent period doubling bifurcation giving rise to alternating LH surge amplitudes. Minimal LH levels remain relatively constant. This suggests that the reduced framework responds to elevated T by altering the amplitudes and timing of LH surges, with sustained oscillations under normal luteinization. Although the dynamic mechanisms governing ultimate dysfunction may differ from the original model, we are able to capture disruptive behavior, which results in elongated IOIs and a decreased number of ovulatory events within a given time span. Specifically, as $\alpha$ increases, the number of ovulatory cycles per year decreases from a maximum of twelve per year to four per year. 

\begin{figure}[p]
\centering
\includegraphics[width=\textwidth]{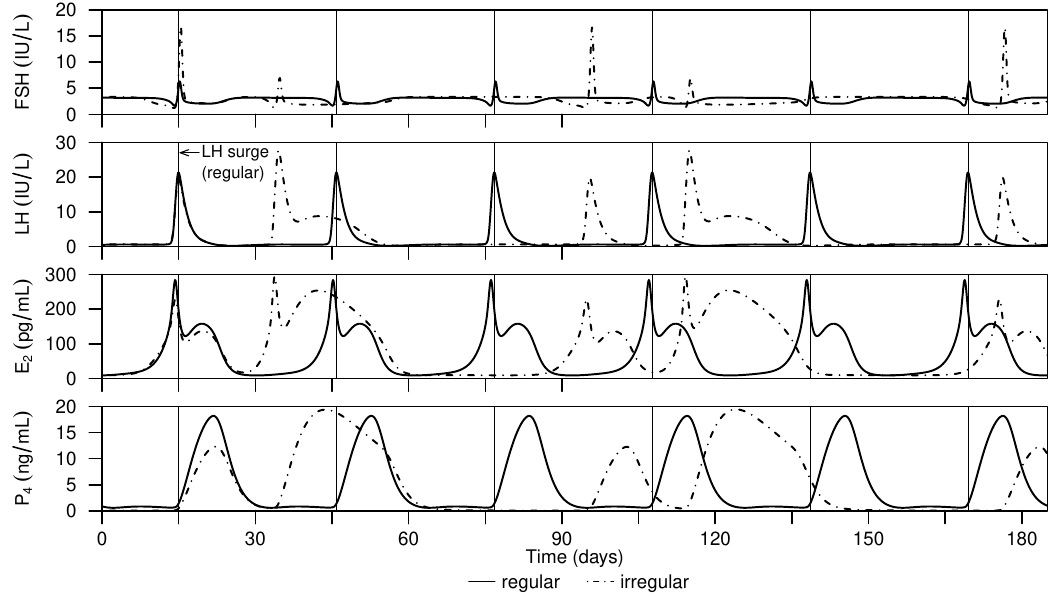}
\caption{Comparison of representative regular and irregular trajectories simulated by the reduced model. The regular cycle displays a characteristic length of 30.9 days. The irregular cycle has a total length of 80.7 days, with IOIs of 19.5 and 61.2 days.}
\label{fig:traj}
\end{figure}
\subsubsection{Phenotype extremes}
From Steps 3 and 4 of Algorithm \ref{algorithm}, we obtain an ensemble of model trajectories. Figure \ref{fig:traj} shows primary hormone trajectories over 186 days for two model solutions, one regular and one irregular, as defined in Step 5\ref{alg:pheno} of the evaluation algorithm. For reference, the timing of the LH surge for the regular phenotype is indicated with a vertical line. Stable limit cycle behavior is exhibited for the regular cycle with a characteristic length of 30.9 days. The irregular phenotype, however, consists of nonuniform behavior of the major hormones. Specifically, the irregular cycle has a length of 80.7 days, with 19.5 and 61.2 days passing between consecutive LH surges. Although hormone levels are relatively normal through the course of the irregular cycle, there are marked differences in hormone patterns that could suggest ovulatory dysfunction.  

\begin{figure}[p]
\centering
\includegraphics[width=.9\textwidth]{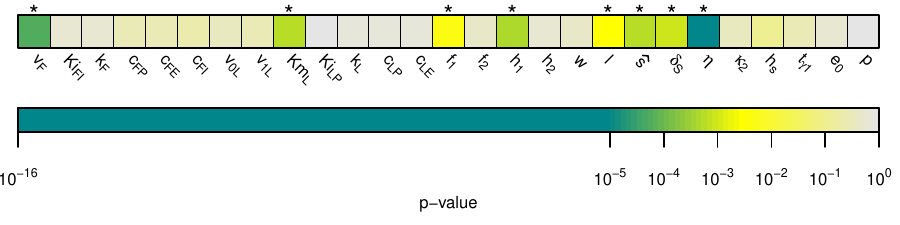}
\caption{Two-sample Kolmogorov-Smirnov  test. Shaded according $p$-value. $^*p<0.01$.}
\label{fig:kstest}
\end{figure}

{\small
\newcounter{params}
\setcounter{params}{0}
\stepcounter{params}
\begin{table}[p]\small
\caption{Eight parameters identified as most important based on the Kolmogorov-Smirnov test. Parameters are ranked in order from most (1) to least (8) significant, according to the $p$-value obtained.}
\label{tab:eightpars}
\centering
\begin{tabular}{clp{1.9in}}\hline
Rank & Name & Description\\\hline
\theparams & $\eta$ &  luteal \etwo\ production; \\\stepcounter{params}\theparams
& $v_F$	 	&   maximal FSH synthesis rate;\\\stepcounter{params}\theparams
& $h_1$		 &  follicle sensitivity to FSH; \\\stepcounter{params}\theparams
& $\hat s$	 & LH support maximal growth rate; \\ \hline
\end{tabular}\!\!
\begin{tabular}{|clp{2in}}\hline
Rank & Name & Description\\\hline
\stepcounter{params}\theparams
& $K_{mL}$	 &  half-maximal \etwo\ stimulation level; \\\stepcounter{params}\theparams
& $\delta_s$  &  LH support decay rate; \\\stepcounter{params}\theparams
& $l$  		& maximal luteolysis rate; \\\stepcounter{params}\theparams
& $f_1$ 	& maximal follicle growth rate. \\ \hline
\end{tabular}
\end{table}
}

\subsubsection{Important parameters: Identification and distributions} \label{sec:pardist}

Using the results from Step 5\ref{alg:pardist} of Algorithm \ref{algorithm}, we can use the Kolmogorov-Smirnov test to assess whether each parameter distribution differs from its counterpart in the opposing phenotype. Results for subsampled parameter distributions are illustrated in Figure \ref{fig:kstest}.  Each box is shaded according to the minimal level of significance that allows us to accept the alternative hypothesis, i.e. that regular and irregular distributions are statistically different. Darker shaded squares correspond to higher levels of significance. Of the 27 parameters remaining in the reduced model, we identify eight that have significantly different distributions between regular and irregular phenotypes, with $p<0.01$ (indicated by $^*$). These parameters are given in Table \ref{tab:eightpars}. Our remaining analysis focuses on these eight important parameters. 

\begin{figure}[bth]
\includegraphics[height=.4\textheight]{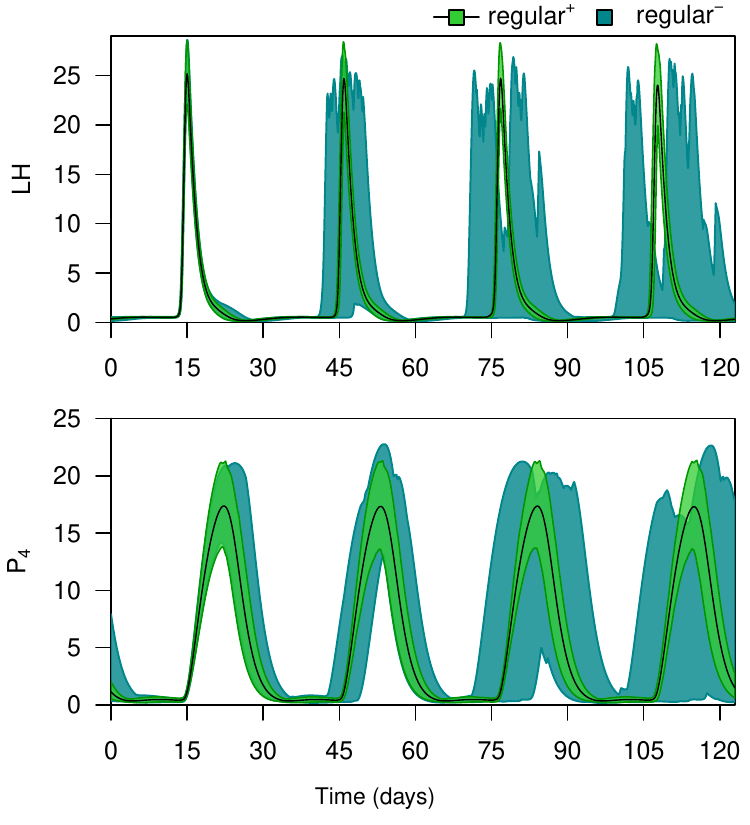}
\hfill
\includegraphics[height=.4\textheight,trim=0.25in 0in 0in 0in,clip=true]{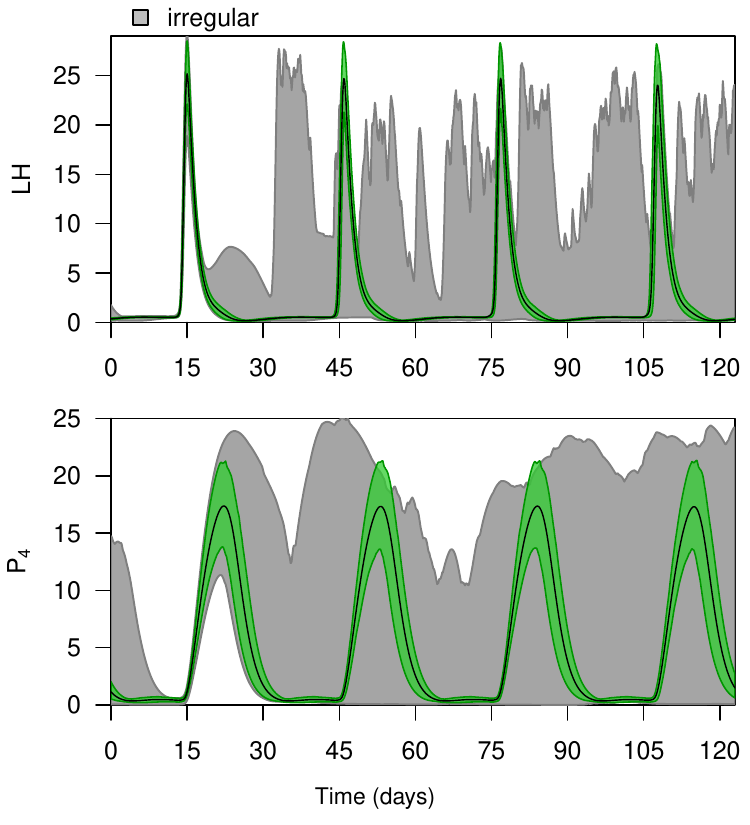}
\caption{95\% confidence intervals of reduced model output over four regular cycles.  Regular$^+$ (green) compared to (\textit{left},teal) regular$^-$ and (\textit{right},gray) irregular phenotypes. Time-dependent regular$^+$ means are indicated with black curves. 
}
\label{fig:hormCI}
\end{figure}
\subsection{Refined phenotypic features}

To determine whether a correlation exists between important parameters and the accuracy of their  accompanying numerical solutions when fit to clinical data, we calculate the mean squared error (MSE) between the model output and the measured data. Although we fail to demonstrate a clear mechanistic relationship between any of the eight relevant parameters (not shown) and their impact on hormone dynamics or phenotypes, we do observe a threshold MSE value---estimated from the MC output---above which all irregular phenotype results lie and below which roughly $85\%$ of regular results lie. We use this threshold to assign an additional subcategory to simulations belonging to the regular phenotype. {Specifically, regular solutions that yield MSE values below the computed threshold, and hence fit hormone data relatively well, are denoted $regular^+$.  Regular solutions that yield above-threshold MSE values, and hence fit hormone data less well, are denoted $regular^-$.  Qualitatively, we consider the regular$^+$ phenotype to reflect `regular IOI-regular dynamics' and regular$^-$ to reflect `regular IOI-irregular dynamics'.}

It is important to recall that there does exist a subset of parameters for which the IOI varies by $50 \%$, where both regular and irregular  IOIs are observed yet the limit cycle length is fixed. Because of this, regular$^+$ implies both low intra-cycle hormone variability compared with data and also low IOI variability.

\begin{figure*}[bth]
\centering
\includegraphics[width=.725\textwidth]{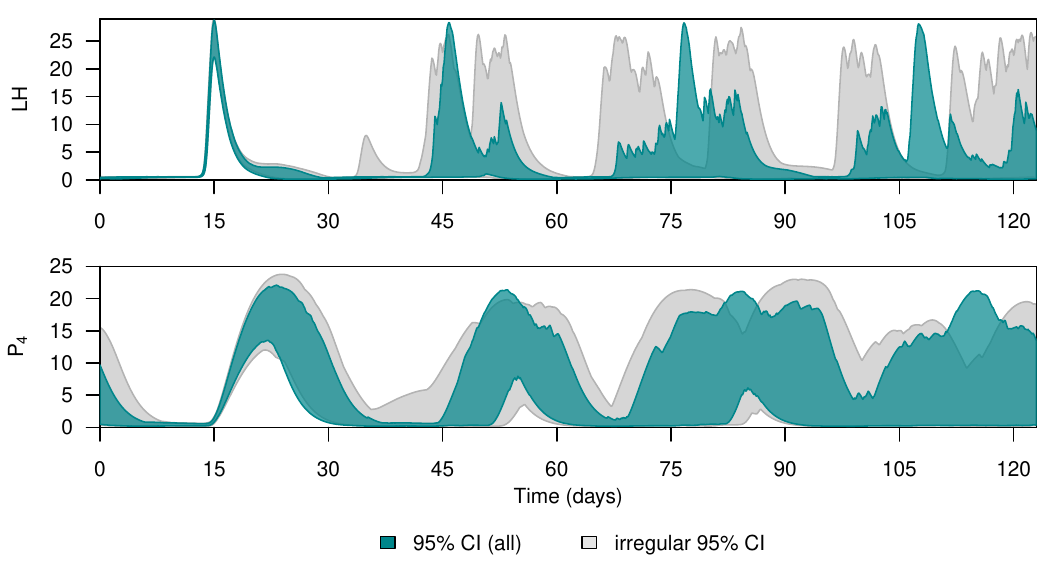}
\caption{95\% confidence intervals (teal) of LH and \pfour\ trajectories satisfying the criterion of \textit{at least one} inter-ovulatory interval of 30--32 days. Light gray: 95\% confidence interval for irregular phenotypes satisfying IOI criterion. 
}
\label{fig:CI30}
\end{figure*}
\begin{figure*}[bth]
\centering
\includegraphics[width=0.6\textwidth]{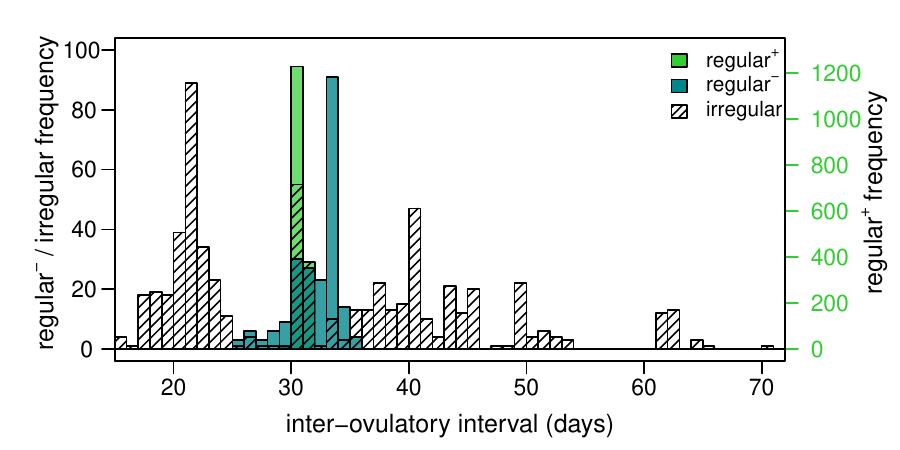}
\caption{
Distribution of inter-ovulatory intervals (IOIs) across phenotypes. Histogram computes the range of frequencies based on individual IOIs, rather than the set of IOIs belonging to independent trajectories. Irregular phenotypes exhibit significantly more variation in IOI than regular phenotypes.}
\label{fig:IOIdistribution}
\end{figure*}
\begin{figure*}[thb]
\centering
\includegraphics[width=0.35\textwidth,trim=0in 0in 0in 0.11in,clip=true]{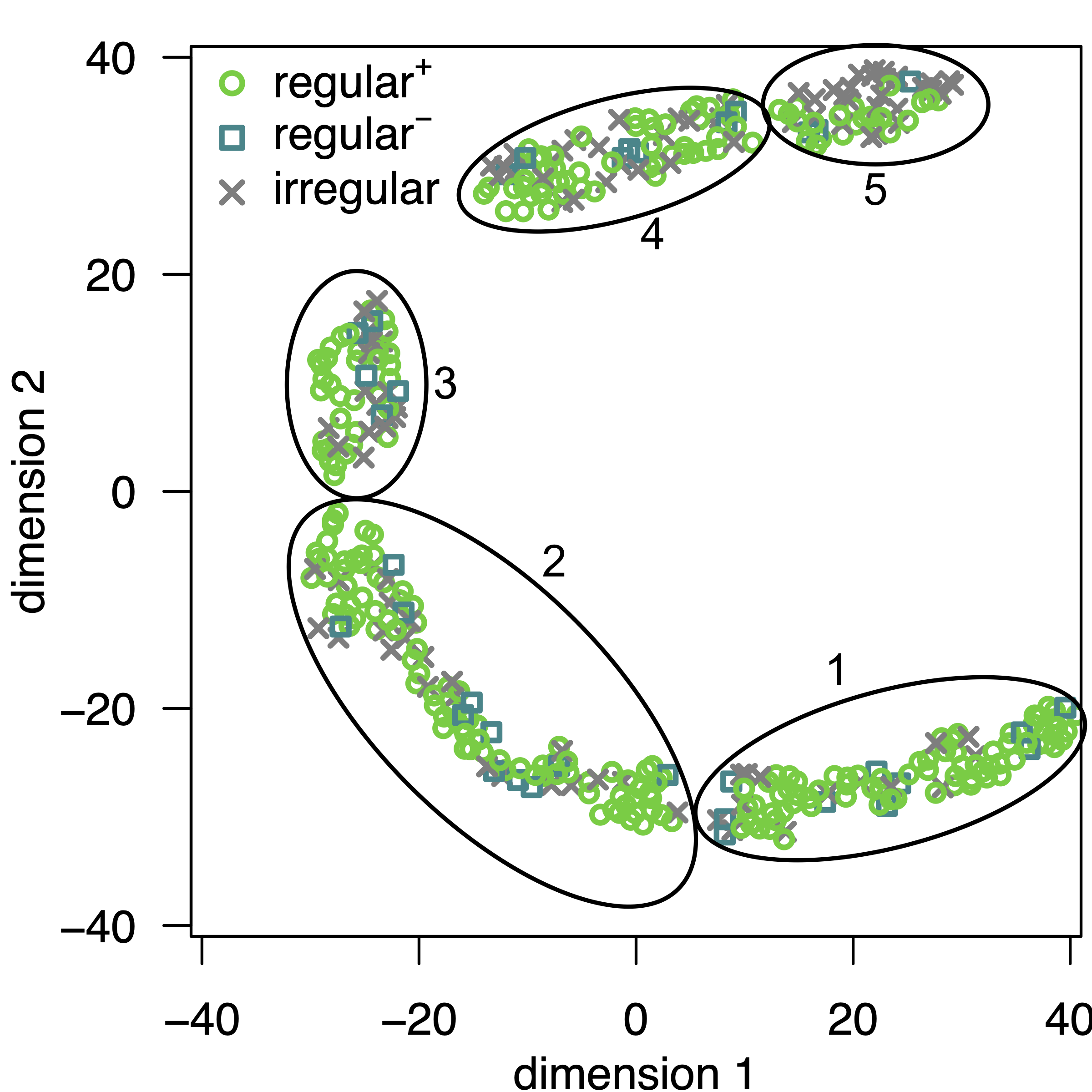}\quad
\caption{$t$-Distributed Stochastic Neighbor Embedding of model results. Dimensional reduction of identified phenotypes based on the eight significant parameters $\eta, ~v_F,~ h_1,~ \hat s,~ K_{mL},~ \delta_s, ~l,~f_1$ gives a two-dimensional embedding of model output.}
\label{fig:tsne}
\end{figure*}

In Figure \ref{fig:hormCI}, we compute $95\%$ confidence intervals of simulated hormone concentrations over four months to examine how hormone profiles influence these refined phenotypes.  As before, we align the simulated LH surge of the first cycle at day 15. Regular$^+$ simulations exhibit the least variation across all cycles (green regions).  Beyond the first LH surge, regular$^-$ phenotypes (left panel, teal regions) have more variation in the timing of characteristic ovulatory events (e.g. LH surge and luteal formation) than regular$^+$, but considerably less variation than the irregular phenotypes (right panel, gray region). As a result, predictability of ovulation is reduced when we refine phenotypes according to data fitting.

On the other hand, if we are less strict with our definition of `normal', we find that it is more difficult to discern reproductive phenotypes. In Figure 
\ref{fig:CI30}, we plot the $95\%$ confidence intervals for all LH and \pfour\ trajectories satisfying the criterion that \textit{at least one} IOI is between 30 and 32 days long (teal). For comparison, we also include the $95\%$ CI for applicable irregular trajectories (gray). Limited to information on a single IOI, there is considerable overlap between opposing phenotypes, which may obscure our ability to discern irregularities in hormone regulation. {These results are important because they highlight how \emph{insufficient data} can both mask ovulatory dysfunction and obscure phenotype definition, discovery, and analysis.} 

In Figure \ref{fig:IOIdistribution}, we examine the distribution of IOIs for each phenotype. Frequencies are determined by the collection of all IOIs, rather than a statistic describing generalized behavior. This is especially useful for the irregular case, which displays much wider variability than either of the regular phenotypes. Further, there appear to be multiple modes in the distribution of IOIs for irregular trajectories, observed at IOIs of 20, 30, and 40 days.  In terms of mathematically versus clinically cyclic behavior, we find that although most simulations result in oscillations, some do not exhibit limit cycle behavior with a characteristic IOI over the 6 months simulated.

\subsection{Dimensional reduction of phenotypes}
To examine refined phenotypes based on parameter estimates, we implement a t-SNE of the parameter profiles, with points distinguished according to the assigned primary and secondary phenotypes. We again limit our analysis to the eight significant parameters found in Section \ref{sec:pardist}. In a two-dimensional reduction of the eight-dimensional parameter space, we find no discernible differences between phenotypes. Instead, five clusters do emerge from the two-dimensional t-SNE, which have been arbitrarily numbered one through five in Figure \ref{fig:tsne}. These results indicate that the set of significant parameters cannot alone isolate reproductive phenotypes.

\begin{figure*}[bth]
\centering
\includegraphics[width=.9\textwidth]{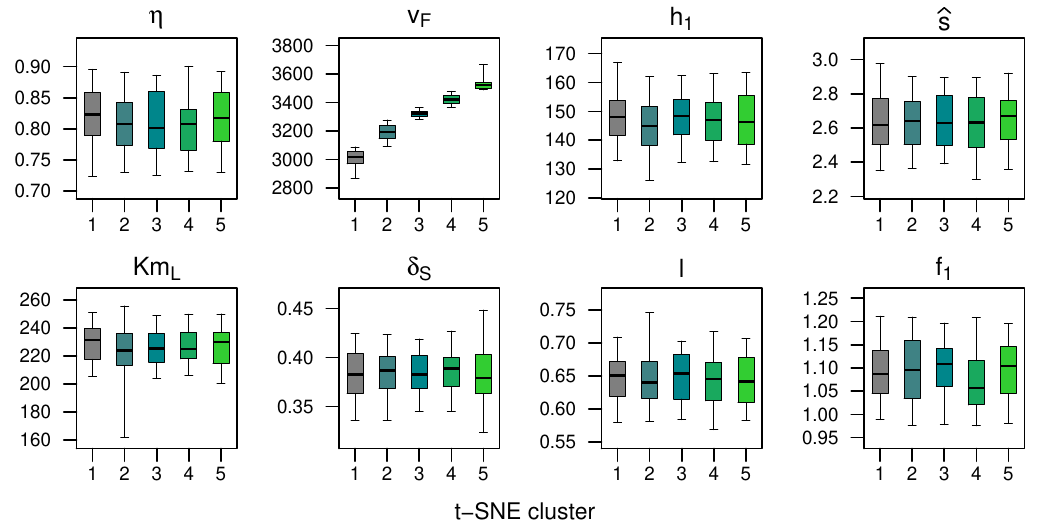}
\caption{Significant parameter estimates for t-SNE clusters.  Cluster-specific behavior is evident for parameter $v_F$, which corresponds to the maximal rate of FSH synthesis in the brain.}
\label{fig:clusters}
\end{figure*}

We explore the characteristics of the five t-SNE clusters further by plotting the individual parameters according to cluster (see Figure \ref{fig:clusters}). Of the eight important parameters we have identified, $v_F$---representing the maximal rate of FSH synthesis---is the only one that exhibits clear cluster-specific behavior. The other parameters vary by group, but not in any clearly discernible way. In Table \ref{tab:clusters}, we   calculate the distribution of regular and irregular phenotypes present in each cluster, accompanied by the mean $v_F$ attained within each grouping. We find that $v_F$ is positively correlated with irregular phenotypes, to the extent that lower values of $v_F$ occur with more regular ovulatory cycles. This suggests that the reduced model introduced herein displays ovulatory dysfunction as a by-product of elevated FSH production. 

\begin{table}[bth]
\centering 
\begin{tabular}{lccccc}
 & \multicolumn{5}{c}{\bf t-SNE Cluster}\\
    &        1  &   2   &  3   &  4  &  5\\\hline\hline
\it regular$^{\pm}$ & 0.86 & 0.81 & 0.70 & 0.71 & 0.48 	\\ 
\it irregular & 0.14	& 0.19 & 0.30 & 0.29 & 0.52 \\ \hline
mean $v_F$ & 3008.0 & 3189.1 & 3321.8  & 3421.8 & 3530.7
\end{tabular}
\caption{Proportion of individual trajectories that qualify as regular or irregular, as distinguished by the t-SNE clusters. Note: \textit{regular}$^{\pm}$ phenotype is the sum of regular$^+$ and regular$^-$ proportions.}
\label{tab:clusters}
\end{table}

\section{Discussion}\label{sec:discuss}

We introduce a new, reduced endocrine model that inherently demonstrates both regular and irregular phenotypes classified by the timing of ovulation. The model produces distinct phenotypes as a result of altered time-independent parameter regimes and in the absence of disease-specific factors, e.g. testosterone-mediated dysfunction in PCOS. Through a comprehensive model evaluation algorithm, we identify a subset of model parameters that provide insight into physiological mechanisms of dysfunction. Further, the reduced framework provides a testable hypothesis of model prediction: consistently similar inter-ovulatory intervals (IOIs) between individuals likely reflect similar reproductive hormone dynamics. But, such consistency is a limiting factor in our ability to broadly predict ovulatory function and highlights the fact that a small set of parameters can produce large variations in ovulatory profiles. These results also imply that there is potentially a many-to-one relationship between endocrine states, e.g., physiologic parameters, and observable endocrine dynamics and dysfunction, e.g., hormone dynamics. This fuzzy causation is not uncommon in physiologic systems or in biomedicine broadly; but to develop better clinical treatment, it is critical to minimize the number of potential causes of an observable problem while maximizing the understanding of the physiologic mechanics driving endocrine dynamics. While we further clarify these issues below---we identify potentially testable mechanisms that drive different endocrine dynamics and phenotypes---substantial problems remain.

Based on the most significant parameters identified by the present work, the model highlights mechanisms associated with pituitary hormone synthesis ($v_F$, $K_{mL}$), follicle growth ($h_1$, $f_1$), luteal dynamics ($\hat s$, $\delta_s$, $l$), and ovarian \etwo\ production ($\eta$). However, the redundancy in the biological processes associated with these parameters allows us to more succinctly characterize sources of dysfunction based on two major processes: altered follicular growth and feedback associated with \etwo\ concentrations. 

\textit{In vitro} experiments suggest that granulosa cells may be more sensitive to FSH in PCOS, affecting follicle growth \cite{YenJaffe}.  Follicular growth is stimulated by FSH, and the model's maximal FSH synthesis rate parameter modulates pituitary stores of FSH. In the irregular phenotype, there is a tendency toward increased mid-cycle FSH levels, which are considered elevated for physiological FSH concentrations (roughly 20 IU/L). In addition, increased $v_F$---identified as a distinguishing parameter in our t-SNE analysis---accompanies increased peak FSH levels, regardless of phenotype. This suggests that the reduced model accounts for ovulatory disruption through changes in FSH, which is also consistent with the current literature, wherein elevated FSH is a determining factor in premature ovarian insufficiency (POI) \cite{Mikhael2019,Jiao2021}. Although the maximal FSH levels produced by the model are relatively lower than those expected from a confirmed POI individual, these levels also occur in the face of residual ovulatory function, albeit irregular. 

Variations in \etwo\ are implicated in multiple manifestations of ovulatory dysfunction. For example, decreased \etwo\ is characteristic of menopausal women. Prolonged exposure to elevated \etwo\ has been associated with ovulatory disruption in previous mathematical models \cite{Clark2003, Harris2014}, and elevated \etwo\ formation has been found in \textit{in vitro} PCOS models \cite{YenJaffe}. In addition, in the current work, parameters associated with luteal stage dynamics are altered in the irregular phenotype. In particular, appearance and disappearance rates of LH support are increased and decreased, respectively. This supports greater ovarian mass during the luteal phase, which contributes to significantly elevated \etwo\ during this period.  Simulated irregular cycles are also associated with higher \etwo\ production rates from functional luteal cells and increased pituitary sensitivity to \etwo, which can prematurely trigger the LH surge. Elevated subthreshold \etwo\ prolongs suppression of FSH and LH release into the serum, thereby inhibiting follicle growth. In extreme cases, this results in two ovulation events close together, followed by an increased period of ovulatory suppression. This is exhibited in Figure \ref{fig:traj}, with a two-month lapse between ovulation events in the representative irregular phenotype. 

The reduced framework is amenable to modifications allowing us to explore testosterone-mediated ovulatory dysfunction. Clinically, it remains unclear how disruptions propagate in the face of hyperandrogenism. We find that when we alter pituitary-specific processes---particularly with respect to LH production---and follicle growth processes with linearly increasing levels of T, cyclic behavior ceases. Further, the steady state approached for sufficiently large insulin influence includes a clinically low level of LH.  In contrast, LH is often found to be elevated in PCOS populations, but with high interindividual variability. These results suggest that we may not associate the T-mediated disruptions within the reduced framework with specific PCOS symptoms, but rather as part of a more generalized manifestation of ovulatory dysfunction due to abnormal responses in the pituitary-ovarian axis.

In the absence of testosterone-mediated modifications, all phenotypes in the new endocrine model exhibit successful ovulatory events, which may vary in frequency.  Further, the  hormone concentrations arising from irregular cycles lie within their respective physiological ranges. Interestingly, the range of IOI for irregular phenotypes is consistent with the ranges reported for individuals near menarche or approaching menopause \cite{YenJaffe}.   The model cannot, nor is it designed to, produce an increase in small ovarian cysts that can accompany PCOS. Yet, it does capture observable information---such as cycle length and the absence of androgen excess---that could indicate a less severe phenotype of PCOS, which would be characterized mathematically by oligo-ovulation. It also appears that our ability to identify defects via reproductive hormones depends on the sampling frequency of data.

The over-arching goal is to use models for predictive decision support and to deepen our understanding of physiology. We wish to not only understand mechanisms of function but also the factors that differentiate those mechanisms. Endometriosis and PCOS are two high-impact  disorders  governed by physiology, both with incompletely understood etiologies. We wish to shed insight on these disorders to better inform intervention and treatment decisions. The current model and evaluation process allows us to delineate dysfunction based on physiology. As constructed, the model is flexible enough to allow us to highlight important---generalizable {or} disorder-specific---mechanisms of dysfunction. What is holding our understanding back now is availability of data.

\section*{Declarations of interest}
None.

\section*{Funding}
E.J.G. reports funding from the Simons Foundation [MPS 585858]. D.A. reports funding from the NLM R01 [LM012734].


\bibliography{ENDOreferences}


\appendix


\section{Original Graham-Selgrade Model Equations \cite{Graham2017}} \label{appendix:orig}
{\small
\begin{flalign}
& \text{Releasable FSH: } &&
	\deriv{FSH_{\rho}}{t} = \frac{v_F}{1+c_{F,I}\frac{S\Lambda}{K_{iF,I} + S\Lambda}} - k_{F} \frac{1+c_{F,P}P_4}{1+c_{F,E}E_2^2}FSH_{\rho} \label{eq:rfshOG}&\\
&\text{Serum FSH: }&&
	\deriv{FSH}{t} = \frac{1}{V} \cdot k_{F} \frac{1+c_{F,P}P_4}{1+c_{F,E}E_2^2}FSH_{\rho} - \delta_{F} FSH \label{eq:fshOG}&\\
&\text{Releasable LH: }&&
	\deriv{LH_{\rho}}{t} = \left[ \frac{v_{0L}  T}{K_{L,T}+T}
	 +  \frac{v_{1L}E_2^n}{K_{mL}^n + E_2^n}\right] \cdot \label{eq:rlhOG} 
	 \frac{1}{1+\frac{P_4}{K_{iL,P}\left(1+c_{L,T}T\right)}}  -  k_{L} \frac{1+c_{L,P} P_4}{1+c_{L,E}E_2}LH_{\rho}   &\\
&\text{Serum LH: }&&
	\deriv{LH}{t} = \frac{1}{V} \cdot k_{L} \frac{1+c_{L,P} P_4}{1+c_{L,E}E_2}LH_{\rho} - \delta_{L} LH \label{eq:lhOG} &\\
&\text{ {Follicular phase:}}&&
\deriv{\Phi}{t} =  f_0 \cdot \frac{T}{T_0}
	+  \left[\frac{f_{1} FSH^2}{\left(\frac{h_{1}}{1+c_{\Phi,T}T/T_0}\right)^2 +FSH^2} 
	 -	\frac{f_2LH^2}{\left(\frac{h_{2}}{1+c_{\Phi,F}FSH}\right)^2 + LH^2} \right] \cdot \Phi    \label{eq:phiOG}\\
&\text{ {Ovulatory phase:}}&&
\deriv{\Omega}{t} =
		\frac{f_2LH^2}{\left(\frac{h_{2}}{1+c_{\Phi,F}FSH}\right)^2 + LH^2} \cdot \Phi - 
		w   S   \Omega \label{eq:omegaOG} \\
&\text{ {Luteal phase: }}&&
\deriv{\Lambda}{t} = w S  \Omega - l  (1-S)\Lambda  
  \label{eq:lambdaOG}\\
&\text{ {LH Support: }}&&
\deriv{S}{t} =  \frac{\hat sLH^m}{h_s^m + LH^m}\cdot(1-S) - \delta_s S \label{eq:sOG}\\
&\text{ {Serum T: }}&&
\deriv{T}{t} = t_0 - \delta_T T  
		+ \left[t_1\mathcal G_1 \left(F_1 + c_{T,F_2}F_2\right)  
		+ t_2\mathcal G_1\mathcal G_2 F_1 \right]
		\cdot \label{eq:tbOG} \\
		&&& \qquad \cdot	{\left[\Phi + \tau_1\Omega + \tau_2 S\Lambda + \tau_3 \left(1-\frac{\Phi+\Omega+\Lambda}{\Psi}\right)\right]}  \notag \\
&\text{ {Intermediate T:}\,\,\,}&&
\deriv{T_\gamma}{t} = t_{g1} \mathcal G_1\mathcal G_2F_1 - \frac{t_{g2}FSH}{h_3+FSH}  T_\gamma  \label{eq:tgOG}\\
&\text{ {Serum E$_2$: }}&&
\deriv{E_2}{t} = e_0 - \delta_E E_2 + \frac{t_{g2}FSH}{h_3+FSH} T_\gamma   \cdot 
		{[\Phi + \eta \Lambda  S]} \label{eq:e2OG}\\
&\text{ {Serum P$_4$: }}&&
\deriv{P_4}{t} =  - \delta_P P_4  + \frac{pLH}{LH+h_p}  \cdot{ \Lambda  S  } \label{eq:p4OG}
\end{flalign}
}
\paragraph{Functional Forms}~\\
\begin{minipage}[t]{.6\textwidth}\vspace{0pt}
$\bullet$ \textbf{Insulin-stimulated conditions ($\alpha > 0$)}\\
     $
    	\begin{aligned}
			\quad&\mathcal G_1 = \mathcal G_1(\alpha)\\
			&\mathcal G_2= \mathcal G_2(\alpha)\\
            &\mathcal D (\alpha) = LH^2 \left[\mathcal G_2 + A\right] + LH 
            	\left[\mathcal G_2 B + A\cdot \left(B+ C\right) \right]+ A\cdot{B\cdot C}\\
            &F_1(LH,\alpha) = LH^2/\mathcal D(\alpha)\\
            &F_2(LH,\alpha) = LH/\mathcal D(\alpha)
        \end{aligned}
    $
\end{minipage}\qquad
\begin{minipage}[t]{.35\textwidth}\vspace{0pt}
$\bullet$ \textbf{Basal conditions ($\alpha = 0$)}\\
     $
     \begin{aligned}
     	\quad&\mathcal G_1 = \mathcal G_2 = 1\\
        &\kappa_1 = 1+ A\\
        &\kappa_2 = B + A(B+C)\\
        &\kappa_3 = ABC\\
        &\mathcal D = \kappa_1 LH^2  + \kappa_2 LH +\kappa_3\\
        &F_1(LH) = LH^2/\mathcal D\\
        &F_2(LH) = LH/\mathcal D
    \end{aligned}
    $
\end{minipage}

\section{Reduced Model Parameters}
\begin{table}[H]
\centering
\begin{tabular}[t]{ll}
\multicolumn{2}{c}{Pituitary Parameters}\\[.5em]
\hline \bf Parameter & \bf Value \\
\hline
 $v_{F}$	&	3219.9	\\
    $K_{iF,I}$	&	149.76	\\
    $k_{F}$	&	3.0212	\\
    $c_{F,P}$	&	65.229	\\
    $c_{F,E}$	&	0.0024047	\\
    $c_{F,I}$	&	3.0188	\\
    $v_{0L}$	&	308.35	\\
    $v_{1L}$	&	44700	\\
    $Km_{L}$	&	226.37	\\
    $K_{iL,P}$	&	3.2279	\\
    $k_{L}$	&	0.67146	\\
    $c_{L,P}$	&	0.015844	\\
    $c_{L,E}$	&	0.00068867	\\ \\
         \end{tabular}\qquad
    \begin{tabular}[t]{ll}
\multicolumn{2}{c}{Ovarian Parameters}\\[.5em]
\hline \bf Parameter & \bf Value \\
\hline
    $f_{1}$	&	1.0958	\\
    $f_{2}$	&	46.225	\\
    $h_{1}$	&	146.31	\\
    $h_{2}$	&	798.39	\\
    $w$	&	0.23497	\\
    $l$	&	0.64178	\\
    $\hat s$	&	2.6338	\\
    $\delta_{S}$	&	0.38256	\\
    $\eta$	&	0.81426	\\
    $\kappa_{2}$	&	8.276	\\
    $h_{s}$	&	11.691	\\
    $t_{g1}$	&	6.3594	\\
    $e_{0}$	&	9.6377	\\
    $p$	&	0.22851	\\ 
\end{tabular}
\caption{Optimal parameters generated from fitting the reduced model to original model output. Listed here are the estimated parameters. Other fixed parameters appearing in the model remain unchanged from \cite{Graham2017}.}
\label{tab:parREDUCED}
\end{table}

\section{Derivation of Testosterone-Dependent Terms} \label{appendix:tst}
To incorporate testosterone implicitly in the reduced model, we need to modify parameters $v_{0L},\, K_{iL,P}$, and $h_1$. We will use $\tilde p$ to denote parameters used in the original Graham-Selgrade model, which we will then redefine to incorporate into the reduced framework.

\paragraph{Derivation of $\xi_1$}
In the original model, basal LH synthesis occurs at rate $\tilde v_{0L} {T}/({T+\beta_1})$, where $\beta_1 = K_{L,T} = 420$. We assume for the reduced model that
\[
v_{0L}\xi_1 = \tilde v_{0L} \frac{T_\alpha}{T_\alpha+\beta_1},
\]
where $\tilde v_{0L}$ is redefined so that $\xi_1=1$ when $T_\alpha = T_0$. That is, we define
$
\tilde v_{0L} =  v_{0L} (T_0 + \beta_1)/T_0.
$
It follows that
\[
v_{0L}\xi_1 = v_{0L} \frac{T_0 + \beta_1}{T_0} \frac{T_\alpha}{T_\alpha+\beta_1} = 
v_{0L}   \frac{(\beta_1+T_0) \cdot T_\alpha}{(\beta_1+T_\alpha) \cdot T_0}.
\]

\paragraph{Derivation of $\xi_2$}
In the original model, \pfour\ inhibition of LH synthesis is scaled by the factor $\tilde K_{iL,P}(1+\beta_2 T) $, where $\beta_2 = c_{L,T} = 0.00959$. Similar to the derivation of $\xi_1$, we assume
\[
K_{iL,P}\xi_2 = \tilde K_{iL,P}(1+\beta_2 T_\alpha),
\]
so that
\[
\tilde K_{iL,P} = \frac{K_{iL,P}}{1+\beta_2 T_0}
\quad \text{ and }\quad 
K_{iL,P}\xi_2 = K_{iL,P}\frac{1+\beta_2 T_\alpha}{1+\beta_2 T_0}. 
\]

\paragraph{Derivation of $\xi_3$}
In the original model, follicle sensitivity to FSH has the form $h_1/[1+\beta_3 T/T_0]$, where $\beta_3 = c_{\Phi,T} = 0.19878$. We assume
\[
h_1 \xi_3 = \frac{\tilde h_1}{1+\beta_3 T_\alpha/T_0},
\]
so that 
\[
\tilde h_1 = h_1 (1+\beta_3),
\]
which implies 
\[
h_1 \xi_3 = h_1 \frac{1+\beta_3}{1+\beta_3 T_\alpha/T_0}.
\]


\section{Empirical Distributions by Phenotype}
~
\begin{figure}[h]
\centering
\includegraphics[width=\textwidth]{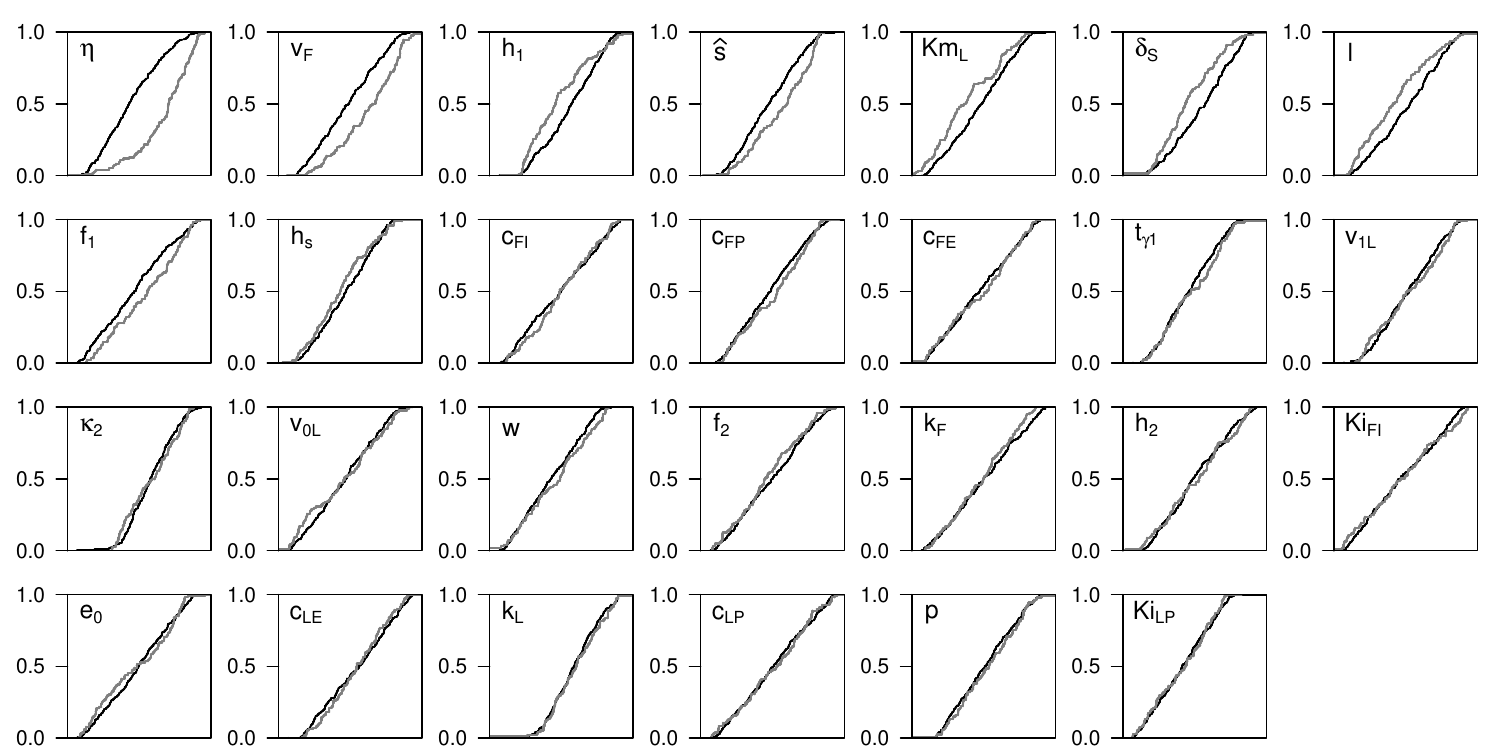}
\caption{Empirical cumulative distribution functions for reduced model parameters, separated by regular (black) and irregular (gray) phenotypes. Parameters are listed, beginning from the top row, in order of decreasing significance.}
\end{figure}


\end{document}